\documentstyle[12pt,epsf]{article}
\newlength{\abstractwidth}
\renewcommand{\thefootnote}{\fnsymbol{footnote}}
\renewcommand{\thanks}[1]{\footnote{#1}} 
\newcommand{\starttext}{
\setcounter{footnote}{0}
\renewcommand{\thefootnote}{\arabic{footnote}}}
\renewcommand{\theequation}{\thesection.\arabic{equation}}
\newcommand{\be}{\begin{equation}}
\newcommand{\bea}{\begin{eqnarray}}
\newcommand{\eea}{\end{eqnarray}}
\newcommand{\beq}{\begin{equation}}
\newcommand{\ee}{\end{equation}}
\newcommand{\eeq}{\end{equation}}

\def\ba{\begin{eqnarray}}
\def\ea{\end{eqnarray}}

\def\12{{1 \over 2}}

\def\dm{dimensional}
\def\3fm{three-form potential}
\def\bs{bosonic string theory}

\def\sz{$S^1/Z_2$}
\def\cc{cosmological constant}
\def\bm{bosonic M theory}
\def\[{\left [}
\def\]{\right]}
\def\({\left (}
\def\){\right)}
\begin{document}
\renewcommand{\theequation}{\thesection.\arabic{equation}}
\begin{titlepage}
\bigskip
\rightline{} \rightline{SU-ITP 00-14} \rightline{hep-th/0012037}
\bigskip\bigskip\bigskip\bigskip
\centerline{\Large \bf {Bosonic M Theory }}
\bigskip\bigskip
\bigskip\bigskip

\centerline{\it Gary T. Horowitz${}^1$ and Leonard Susskind${}^2$}
\bigskip
\centerline{${}^1$Department of Physics} \centerline{University of
California, Santa Barbara} \centerline{Santa Barbara, CA. 93106}
\medskip
\centerline{${}^2$Department of Physics} \centerline{Stanford
University} \centerline{Stanford, CA 94305-4060}
\bigskip\bigskip


\begin{abstract}
We conjecture that there exists a strong  coupling limit of
bosonic string theory which is related  to the 26 dimensional
theory in the same way that 11 dimensional M theory is related to
superstring theory. More precisely, we believe that bosonic string
theory is the compactification on a line interval of a 27
dimensional theory whose low energy limit contains gravity and a
three-form potential. The line interval becomes infinite in the
strong coupling limit, and this may provide a stable ground state
of the theory. We discuss some of the consequences of this
conjecture.

\medskip
\noindent
\end{abstract}
\end{titlepage}
\starttext \baselineskip=18pt \setcounter{footnote}{0}

\setcounter{equation}{0}
\section{Introduction}

There is growing evidence that all the known perturbative ten
dimensional superstring theories are limits of an eleven
dimensional theory called M theory \cite{duff, hull, witten1}. In
particular, the ubiquitous dilaton which controls the string
coupling is simply related to the size of the extra dimension. The
26 \dm\ bosonic string has not been included in these developments mostly
due to the widespread belief that the existence of the tachyon
indicates that the theory is ill defined\footnote{The strong coupling
limits of the ten dimensional bosonic Type 0 theories have been discussed
in \cite{bega}, and other nonsupersymmetric theories have been considered
in \cite{bldi}.}. However during the past
year the significance of the open string tachyon has been
understood. Rather than indicating that the theory is sick, it
just shows that the usual vacuum is unstable. As Sen first proposed \cite{sen1},
this vacuum can be
viewed as a closed string vacuum together with an unstable D
25-brane. There is increasing evidence that there is a stable
minimum of the open string tachyon potential at a value equal to
minus the tension of a D 25-brane, and about this minimum there
are no open string excitations. This raises the
possibility that the closed string tachyon can similarly be
removed by appropriately shifting to a new ground state. However, there are
good arguments that the closed string tachyon cannot be removed by
direct analogy to the open string case. There is probably no
stable minimum of the closed string tachyon potential
\cite{banks}. Something more dramatic is needed.

In this paper we study the strong coupling limit of bosonic string
theory and argue that the tachyon instability may be removed in
this limit. Unfortunately we have very little firm ground to stand
on when trying to determine the strong coupling limit of a theory
without supersymmetry. In this paper we make a guess based on the
assumption that \bs \ is not wholly dissimilar to IIA and
heterotic string theory.

The main clue motivating our guess  comes from the existence of
the dilaton and its connection to the coupling constant. The
action for the massless sector of \bs \ is \be\label{bst} S = \int
d^{26}x \sqrt{-g} e^{-2\phi} \[R + 4\nabla_\mu \phi \nabla^\mu
\phi -{1\over 12} H_{\mu\nu\rho} H^{\mu\nu\rho} \] \ee
 Evidently, as in IIA string
theory, the dilaton enters the action just as it would if it
represented the compactification scale of a Kaluza Klein theory.
We propose to take this seriously and try to interpret \bs \ as a
compactification of a 27 \dm \ theory. We will refer to this
theory as \bm . (The possibility that the bosonic string has a
27 \dm\ origin was also briefly discussed in \cite{rey}, in the
context of a proposed matrix string formulation.)

In the case of IIA string theory there was a second clue that led
to its interpretation as a Kaluza-Klein  compactification of an 11
\dm \ theory; the existence of a vector boson in the string
spectrum. Closed \bs \ does not have a massless vector. This means
it cannot be a compactification on an $S^1$. In this respect the
situation is more analogous to that of heterotic string theory.
 The solution to the strong coupling problem
in that case is a compactification of M theory on a line interval,
or more exactly on the orbifold $S^1/Z_2$ \cite{horavawitten}.
The absence of a $U(1)$
symmetry means that there is no massless gauge boson.  Accordingly,
we propose that closed \bs \ is a compactification of  27 \dm \
\bm \ on $S^1/Z_2$. In the supersymmetric case, to cancel anomalies
one had to add $E_8$ gauge fields at each end of the line interval. In the
bosonic case, since there are no fermions or chiral bosons, there are no
anomalies to cancel. So there are no extra degrees of freedom living at 
the fixed points.

 As in the case of the M theory - heterotic  connection, the weakly
coupled string theory is the limit in which the compactification
length scale becomes much smaller than the 27 \dm \ Planck length
and the strong coupling limit is the decompactification limit. The
27 \dm \ theory should contain membranes but no strings, and would
not have a dilaton or variable coupling strength. The usual
bosonic string corresponds to a membrane stretched across the
compactification interval.

Some support for bosonic M theory comes 
from the following simple observation. The left moving modes of the heterotic
string are precisely those of  the 26 \dm\ bosonic string. It has been argued
\cite{beva} that at least perturbatively,
the right moving modes of the bosonic string can be
embedded in the right moving modes of the heterotic string. So the
 entire bosonic string is contained in the heterotic string. Since we now
know that nonperturbatively the heterotic string grows an extra dimension,
it is plausible that the bosonic string will similarly gain an extra dimension
at strong coupling.

\setcounter{equation}{0}
\section{The Low Energy Theory}

In this section we study the low energy limit of bosonic M theory
which is a gravity theory in 27 dimensions. Without the powerful
tool of supersymmetry it is difficult to give rigorous arguments.
Nevertheless there are some plausible guesses that we can make
about the  form of the low energy action, using the fact that it
must reduce to the usual bosonic string theory in the weak
coupling limit. After deriving the low energy action, we show how
the tachyon instability may be removed in the strong coupling
limit, and then study branes in this theory.

\subsection{Motivation from Weak Coupling Limit}

In order to reproduce the known spectrum of weakly coupled \bs, 
\bm \  will have to contain an additional field besides
the 27 \dm \ gravitational field, namely a three-form potential
$C_{\mu\nu\rho}$. Let us consider the various massless fields that
would survive in the weak coupling limit. First of all, there
would be the 26 \dm \ graviton. As usual, general covariance in 26
dimensions would insure that it remains massless.
The component of the 27 \dm \ gravitational field $g_{27,27}$ is a
scalar in the 26 \dm \ theory. It is of course the dilaton. No
symmetry protects the mass of the dilaton. In fact we know that at
the one loop level a dilaton potential is generated that lifts the
dilatonic flat direction.
 Why the mass vanishes in the weak coupling limit
is not clear.

Massless vectors have no reason to exist since there is no
translation symmetry of the compactification space. This is
obvious if we think of this space as a line interval. If we think
of it as \sz \ then the two fixed points of the orbifold break the
symmetry.

The three-form gauge field $C_{\mu \nu \sigma}$ gives some
massless fields.
 If one of the indices of the three-form is in the compact $27^{th}$
direction, the resulting 26 \dm \ field is the two-form $B_{\mu
\nu}$ which is well known in \bs . It remains massless due to its gauge
invariance. The components of $C_{\mu \nu \sigma}$ in which all
three components are in the 26 \dm \ subspace give a three-form which
is absent in the usual bosonic string  spectrum. Once again we
take a hint from heterotic string theory. In that case the  the
three-form that would be inherited from the 11 \dm \ origin of
heterotic string theory is projected away by the $Z_2$
identification. This is because M theory includes a Chern-Simons
term  which implies that the action is invariant under $Z_2$ only
if $C$ is odd under this identification. In the
present case we will also assume that $C$ is odd
under   the $Z_2$.\footnote{If $y$ is the coordinate along the $S^1$,
the fact that the basis vector $\partial/\partial y$ points 
away from the fixed point
$y=0$, means that it also must change sign under the $Z_2$. This means
that the components $C_{y ij}$ are even under $y\rightarrow -y$, while
$C_{ijk}$ are odd (where $i,j,k$ denote all directions other than $y$).}
Given our limited knowledge of the theory, we
do not  know if this is required by a symmetry of the action or not.

We are thus led to the following low energy action for \bm:
\be\label{bmt} S=\int d^{27}x \sqrt{-\hat g}\[R(\hat g)- {1\over
48}F_{\mu\nu\rho\sigma} F^{\mu\nu\rho\sigma}\] \ee where $F=dC$.
To see the relation to (\ref{bst}),  we set
\be\label{redmet}
\widehat {ds}^2 = e^{2\sigma} dy^2 + e^{-\sigma} g_{\mu\nu} dx^\mu
dx^\nu \ee 
where $\sigma $ and $g_{\mu\nu}$ are functions of $x^\mu$ but
independent of $y$, and set
\be
H_{\mu\nu\rho} = F_{y \mu\nu\rho} \ee The coordinate $y$ takes
values $-1 \le y \le 1$ and we identify $y$ with $-y$.
This prevents a term like $A_\mu dy dx^\mu$ from appearing in
(\ref{redmet}). 
Substituting into the action and integrating by parts yields
\be
S = \int d^{26}x \sqrt{-g} e^{-11\sigma} \[R(g) + 125 \nabla_\mu
\sigma
 \nabla^\mu \sigma - {1\over 12} H_{\mu\nu\rho} H^{\mu\nu\rho}\]
 \ee
There is no four-form in 26 dimensions since, as we have just
explained, it is projected out by the identification on $y$. If we
now define $2\phi \equiv 11\sigma$, this becomes the standard
action for bosonic string theory (\ref{bst}) except that the
coefficient of the $(\nabla \phi)^2$ is off by a factor of
$125/121$. So we recover the right fields and interactions, but
one numerical coefficient is slightly off. This is not a
contradiction since the action (\ref{bmt}) is only valid on scales
larger than the 27 dimensional Planck length, and to recover
(\ref{bst}) we need to take a limit where one direction becomes
much smaller than this. Without supersymmetry to protect
coefficients, they can change as the coupling increases. In this
respect, the factor of  $125/121$ may be analogous to the factor of
$3/4$ which arises in comparing the entropy of weakly coupled
$3+1$ Yang-Mills with the near extremal three-brane \cite{gkp}.

The relation between the 27 \dm\ Planck length $l_p$ and the 26
\dm\ string length $l_s$ and coupling $g=e^\phi$, follows from the
relation between $\sigma$ and $\phi$. Since $g^2 l_s^{24} = G_{26}
= G_{27}/e^\sigma l_p$ we get \be \label{planck}   g^{1/11} l_s =
l_p  \ee Since $g=e^{11\sigma/2}$, weak coupling corresponds to a
small distance in the extra dimension, as expected.

There is a possibility of adding a cosmological constant to the
action (\ref{bmt}). Indeed, in the absence of supersymmetry, it
would appear inevitable that one is generated. We will discuss
this in section 4, but for now, we will assume the cosmological
constant is zero.

\subsection{Tachyon}

We now consider the fate of the closed string tachyon at strong
coupling. The trivial solution to (\ref{bmt}) consisting of $F=0$
and flat spacetime compactified on $S^1/Z_2$ has a nonperturbative
instability. This is analogous to the instability of the
Kaluza-Klein vacuum found by Witten \cite{kkvavuum}, and very similar
to its application  to heterotic -- M theory in 
\cite{horavafabinger}. The process that destabilizes the space is
mediated by an instanton in which the two ends of the world (ends
of the compactification interval) come together and produce a
``hole" in space. In Minkowski space the hole rapidly grows and
eats the entire space.

The appropriate instanton is (a projection of)
the 27 \dm\ euclidean Schwarzschild
metric. \be\label{sch} ds^2 = \[1-\(r_0\over r\)^{24}\] dy^2 +
\[1-\(r_0\over r\)^{24}\]^{-1} dr^2 +r^2 (d\theta^2 + \sin^2 \theta
d\Omega_{24}) \ee The coordinate $y$ is periodic with period $P=
\pi r_0/6$. To apply it to our case, we identify $y$ and $-y$. So
the size of the extra dimension at infinity is $\pi r_0/12$. To
obtain the Lorentzian evolution, one analytically continues $\theta
\rightarrow (\pi/2) + it$. To picture this evolution, consider the
surfaces at the ends of the interval, $y=0, \pi r_0/12$.
The separation between these surfaces goes to zero smoothly at $r=r_0$, so the
two surfaces are really one surface with the shape of a wormhole. 
At the initial time, $t=0$, the  proper size of the wormhole is $r_0$,
but as time evolves, it grows exponentially.

This instanton description is only valid for $r_0\gg l_p$. However
similar instabilities occur in various non--supersymmetric
D--brane systems. A typical example is a D--brane anti D--brane
system. If the distance between
the branes is larger than the string scale, an instanton process
bridging the two branes can again lead to a runaway hole eating
the branes \cite{cama}. When the branes are closer than the string scale the
same process can take place by a perturbative mechanism. At a
critical point the lightest string connecting the branes becomes
massless and then tachyonic \cite{bankssusskind, sen2}. This pattern is seen in several
examples and leads to the following conjecture:

When the two ends of the world are closer than the 27 \dm \ Planck
length a tachyon appears in the spectrum. This is just the closed
string tachyon found in string perturbation theory.

The action for the instanton (\ref{sch}) is proportional to
$(r_0/l_p)^{25}$. So in the limit of strong string coupling, $r_0
\rightarrow \infty$, this nonperturbative instability is
suppressed. Uncompactified 27 \dm\ flat space may be a stable
ground state of bosonic M theory.

\subsection{Branes}

In the absence of supersymmetry, there are no BPS states.
Nevertheless, there are stable brane configurations. In terms of
the low energy action (\ref{bmt}) they arise as the extremal limit
of black brane solutions. Since the charge must be carried by a
four-form, there are 2-branes which are electrically charged and
21-branes which are magnetically charged. It is natural to assume
that there are fundamental 2-branes and 21-branes with Planck
tension, and these black brane solutions describe the
gravitational field of a stack of parallel branes. A nontrivial
check of this idea is to compute the tension of a fundamental
2-brane stretched across the extra dimension. It is given by $T=
e^\sigma/l_p^2$. Using (\ref{planck}), and the relation between
$\sigma$ and $\phi$ we get $T= 1/l_s^2$ which is the right answer
for a fundamental string. Similarly, the tension of a 21-brane which
is not oriented along the extra dimension is $T_{21} = 1/l_p^{22} =
1/(g^2 l_s^{22})$ which again is the right answer for a solitonic
21-brane in string theory.

The black brane solutions can be read off from the general
discussion of nondilatonic black branes in \cite{ght}. The
black 2-brane is given by \bea\label{btb} ds^2 =& -\[1-\(r_+\over
r\)^{22}\]
\[1-\(r_-\over r\)^{22}\]^{-1/3} dt^2
    +\[1-\(r_-\over r\)^{22}\]^{2/3} dx_i dx^i \cr
    &+\[1-\(r_+\over r\)^{22}\]^{-1} \[1-\(r_-\over r\)^{22}\]^{-1} dr^2
    +r^2 d\Omega_{23}
    \eea
with four-form \be    {}^* F = N\ l_p^{22}\ \epsilon_{23}
    \ee
where $\epsilon_{23}$ is the volume form on a unit $S^{23}$. The
charge $N$ is the number of fundamental two branes, and is related
to the two free parameters $r_\pm$ via \be\label{Ntwo} N^2 =
{1100\over 3} \(r_+ r_-\over l_p^2\)^{22} \ee There is an event
horizon at $r=r_+$ and a curvature singularity at $r=r_-$. The
Hawking temperature of this black 2-brane is \be\label{Ttwo}
T={11\over 2\pi r_+}\[1-\(r_-\over r_+\)^{22}\]^{1/3} \ee If we
compactify the two directions along the brane on a torus with side
$L$, then the horizon area is \be\label{Atwo} A= r_+^{23} \
\Omega_{23} \ L^2 \[1-\(r_-\over r_+\)^{22}\]^{2/3} \ee 
where $\Omega_{23}$ is the area of a unit $S^{23}$. In the
extremal limit, $r_+ = r_-$, (\ref{btb}) takes a simpler form by
setting $\rho^{22} = r^{22} - r_-^{22}$: \be\label{exttwo} ds^2 =
f(\rho)^{-2/3}[-dt^2 + dx_i dx^i] + f(\rho)^{1/11}[d\rho^2 +
\rho^2 d\Omega_{23}] \ee where \be f(\rho) = 1 + \(r_-\over
\rho\)^{22} \ee 
This extremal brane has zero Hawking temperature and 
is quantum mechanically stable.
The surface $\rho =0$ is  a smooth horizon. There is no force between
two parallel extremal branes. Static, multi-brane solutions can be obtained
by replacing $f$ with a more general solution of Laplace's equation.

We now turn to the black 21-brane. The metric is \bea ds^2 =&
-\[1-\(r_+\over r\)^{3}\] \[1-\(r_-\over r\)^{3}\]^{-10/11} dt^2
    +\[1-\(r_-\over r\)^{3}\]^{1/11} dx_i dx^i \cr
        &+\[1-\(r_+\over r\)^{3}\]^{-1} \[1-\(r_-\over r\)^{3}\]^{-1} dr^2
     +r^2 d\Omega_{4}
         \eea
and the four-form is $ F= N\ l_p^3\ \epsilon_4 $, where
\be\label{None} N^2 = {75\over 11} \(r_+ r_-\over l_p^2\)^3 \ee
The Hawking temperature is \be\label{Tone} T={3\over 4\pi
r_+}\[1-\(r_-\over r_+\)^{3}\]^{1/22} \ee and the horizon area is
\be\label{Aone} A= r_+^{4} \ \Omega_{4} \ L^{21} \[1-\(r_-\over
r_+\)^{3}\]^{21/22} \ee where we have again compactified the
directions along the brane to have size $L$. Setting $\rho^3=
r^3 - r_-^3$, the extremal limit is
\be\label{extone} ds^2 =  f(\rho)^{-1/11}[-dt^2 + dx_i dx^i] +
f(\rho)^{2/3}[d\rho^2 + \rho^2 d\Omega_{4}] \ee where now \be
f(\rho) = 1 + \(r_-\over \rho\)^{3} \ee Like the 5-brane of M
theory, this 21-brane is completely nonsingular. The spacetime
behind the horizon $\rho=0$ is identical to the spacetime in
front.

We now suppose that one direction of spacetime
is compactified on $S^1/Z_2$, and the four-form $F$ is odd under the
$Z_2$ identification. The situation is similar to the usual heterotic
string construction \cite{llo}. Recall 
that, if $y$ is the coordinate in the compact direction,
the components $F_{yijk}$ must be
even under $y\rightarrow -y$, while
$F_{ijkl}$ are odd (where $i,j,k,l$ denote all directions other than $y$).
Thus, if $y$ is one of the two directions along the 2-brane,
the identification
can be done trivially since the solution 
is invariant. As we have
already noted, this corresponds to $N$ bosonic strings in 26 dimensions. 
If the 2-brane is perpendicular to $y$, a static solution can
still be constructed by putting the 2-brane half way between the two
fixed points and adding an anti-2-brane at its image point under the $Z_2$.
(This solution is not known explicitly and will be unstable.)
It results in an unstable D2-brane
in 26 dimensions.
If the 21-brane is perpendicular to $y$, an invariant solution is obtained
by adding another 21-brane (not anti-brane) at its image point under the $Z_2$.
This
corresponds to a 21-brane  in string theory which is magnetically
charged with respect to the three-form $H$. If $y$ is one of the directions
along the 21-brane, then no invariant solution can be constructed, since
$F_{ijkl} \ne 0$ at $y=0$.

As an aside, we note that there is also a brane solution 
of 26 dimensional bosonic string
theory which has both electric and magnetic charge associated with
the three-form $H$. It is a 21-brane with fundamental strings
lying in it and smeared over the remaining 20 directions.
Dimensionally reducing to six dimensions by compactifying on a
small $T^{20}$, one recovers the usual self dual black string in
six dimensions.

\setcounter{equation}{0}
\section{Holographic Duals}

In this section we go beyond the low energy limit, and try to say
something about exact  \bm. Since it contains gravity it should be
holographic. There are two types of holographic duals that we have
become familiar with. The first is Matrix theory which is based on
the existence of stable D0-branes in type IIA theory and the
existence of a DLCQ quantization of M theory. However in the
present case in which the compactification is on a line interval
rather than a circle this type of construction is questionable 
(but see \cite{rey}).

The other type of holographic dual is through  AdS/CFT duality
\cite{maldacena}. Following the arguments used for the
superstring, we consider the near horizon limit of the extreme
black brane solutions. As usual, the near horizon limit
corresponds to dropping the one in $f$ in the solutions
(\ref{exttwo},\ref{extone}). Starting with the 2-brane, the
resulting space is $AdS_4\times S^{23}$. From (\ref{Ntwo}), the
radius of
  each is proportional to $N^{1/11}$.
The CFT dual would be a 2+1 \dm \ conformal field theory with a
global $SO(24)$ symmetry. The natural candidate would be the \dm \
reduction of 26 \dm \ Yang Mills theory  which has 23 scalars in
the adjoint representation. This theory has manifest $SO(23)$
symmetry. The mechanism for enhancing the symmetry would have to
be similar to the enhancement of $SO(7)$ to $SO(8)$ in the
supersymmetric case. However in the present situation we have no
superconformal symmetry to ensure the enhanced symmetry. A strong
test of the existence of \bm \ is the existence of a conformal
fixed point with  $SO(24)$ symmetry at least in the $N \to \infty$
limit. In other words, if there does not exist a $2+1$ CFT with
$SO(24)$ global symmetry, \bm \ would be disproven.

As in the usual AdS/CFT correspondence, thermodynamics of the CFT
should be related to the near extremal 2-brane.
 From (\ref{Ntwo}) -- (\ref{Atwo}),
 the entropy of near extremal 2-branes can be expressed
\be
S\propto N^{25/22} (LT)^2 \ee This looks like the entropy of a
$2+1$ field theory. The $N$ dependence is analogous to the
$N^{3/2}$ which appears in the usual M 2-brane, and can similarly
be viewed as a  prediction for the density of states of the theory
at strong coupling. 

Since there are also solutions of the form $AdS_4\times K$ where
$K$ is any 23 \dm\ Einstein space, there may also exist holographic duals
of the theory with these boundary conditions. They would be 2+1 conformal
field theories with less symmetry.

Starting with the extreme 21-brane (\ref{extone}), the near
horizon limit is $AdS_{23} \times S^4$, where the radii of each is
proportional to $N^{2/3}$. If the theory exists, its holographic
dual will be a 22 dimensional conformal field theory with a global
$SO(5)$ symmetry. It follows from (\ref{None})--(\ref{Aone}) that
in the near extremal limit, the entropy of the black 21-brane can
be expressed
\be
S\propto N^{25/3} (LT)^{21} \ee Once again, this is consistent
with a 22 dimensional field theory with a large number of degrees
of freedom.

\setcounter{equation}{0}
\section{Discussion}

We have proposed that a bosonic version of M theory exists, which
is a 27 \dm\ theory with 2-branes and 21-branes. One recovers the
usual bosonic string by compactifying on $S^1/Z_2$ and shrinking
its size to zero. In particular, a Planck tension 2-brane
stretched along the compact direction has the right tension to be
a fundamental string. This picture offers a plausible explanation
of the tachyon instability and suggests that uncompactified 27
\dm\ flat space may be stable. A definite prediction of this
theory is the existence of a $2+1$ CFT with $SO(24)$ global
symmetry, which should be its holographic dual for $AdS_4\times
S^{23}$ boundary conditions.

The conjecture that  \bm \ exists raises a number of questions which we
now address:

1) What kind of theory do we get if we compactify \bm \ on a
circle instead of a line interval? Do we get a weakly coupled
string theory in the limit that the circle shrinks to zero?
This seems problematic since,
whatever the resulting theory is, it should have
a massless vector and three-form potential. Of course the open
string has a massless vector, but as far as we know,
there is no 26 \dm\ bosonic string theory with a three-form
potential. Instead we believe the limit of \bm\ compactified on a circle
as the radius $R\rightarrow 0$ is the same as the limit $R\rightarrow \infty$,
i.e., the uncompactified 27 \dm\ theory. If we compactify \bm\ on 
$S^1 \times (S^1/Z_2)$, and take the second factor very small, this is
a consequence of the usual T-duality of the bosonic string. More generally,
it appears to be the only possibility with the right massless spectrum.

2) Must \bm \ have a vanishing \cc ? If not, what is the sign of
the \cc ? If it is negative then there should be a 27 \dm \ AdS
solution. The holographic representation of this theory should be
an isolated 26 \dm \  conformal field theory. Since it is likely
that the \cc \ would be of order one in Planck units we would not
expect classical Einstein gravity to be an accurate description.
The best description would be the CFT. If the \cc \ is positive,
how do we make sense out of the theory in de Sitter space? This
would be the first example of a de Sitter solution emerging out of
string theory.

Even with a cosmological constant  $\Lambda$, there are solutions
of the form $AdS_4\times S^{23}$. The only difference is that the
curvature on the two spaces need not be comparable, and are
related to different combinations of the four-form charge $N$ and
$\Lambda$. If $\Lambda >0$, there is a particularly interesting
special case in which the solution is a sphere cross four
dimensional Minkowski spacetime. This may have phenomenological
applications. It is worth emphasizing that this solution exists
for any (positive) cosmological constant, as long as $F$ can be chosen
appropriately. It would certainly be interesting to find a
dynamical mechanism which would require $F$ to cancel $\Lambda$ in
this way. (For a recent discussion of a possible mechanism, see \cite{lambda}.)
In any event, we find it intriguing that four
dimensional spacetimes arise naturally in this theory.

3) Bosonic string theory contains unstable Dp-branes for all $p$.
What are the analog of these in bosonic M theory? It appears that
most of these do not survive the strong coupling limit and do not
exist as new degrees of freedom in 27 dimensions. This is not
surprising since Type II superstring also has unstable Dp-branes
which do not appear to have an analog in M theory. However, some
Dp-branes may remain. We already saw a construction of an unstable D2-brane
in section 2. D0-branes can be identified with modes in the $27^{th}$
direction. In the theory compactified on $S^1/Z_2$,
these modes are unstable since
they bounce off the fixed
points, interact with themselves and decay into radiation in the other
directions.
In the uncompactified limit, they should become stable.

4) Even if 27 \dm\ flat space, $M_{27}$, is a stable vacuum,
one might ask what is the ``ground 
state" of the theory at finite string coupling, or finite compactification
size? Tachyon condensation is not likely to lead back to $M_{27}$, and there
is probably no stable minimum of the tachyon potential in 26 dimensions 
\cite{banks}. Instead,
we believe tachyon condensation may lead to an exotic state with zero metric
$g_{\mu\nu}=0$.
It is an
old idea that quantum gravity may have an essentially topological
phase with no metric. We have argued that the tachyon instability is 
related to nucleation of  ``bubbles of nothing"  which is certainly reminiscent
of zero metric. Further support for this idea comes from some old 
results on the closed string tachyon. Using modular invariance of the
one loop vacuum amplitude, one can relate the existence of a tachyon to 
the asymptotic density of states. It was shown that the tachyon is absent only
if, at high energies, the theory has at most a finite number of 
fields propagating in two
spacetime dimensions \cite{kutasov}.
Similar results were found by studying the theory
near the Hagadorn transition \cite{atick}.
If the theory starts in 26 (or 27) dimensions, the only way to
get down to two dimensions is to have a highly degenerate metric. The most
symmetric state would then be $g_{\mu\nu}=0$,
and two dimensional subspaces might arise as excitations.

This raises an interesting question in string field theory.
Witten's open bosonic
string field theory \cite{wsft} takes the form
\be\label{sft}
S = \int A*QA + {2g\over 3} A*A*A
\ee
where $Q$ is the BRST operator and $*$ is a noncommutative product.
Formally, $\int$ and
$*$ are independent of the metric and other closed string backgrounds 
but $Q$ is not.
Since an interacting theory of open strings must include closed strings, it is
awkward having this explicit background dependence in the action. 
It was shown in \cite{hlrs}
that (\ref{sft}) could be derived from the purely cubic action
\be
S=\int \Phi*\Phi*\Phi
\ee
There is a solution $\Phi_0$ to the equation of motion $\Phi*\Phi=0$
such that expanding
about this solution, $\Phi=\Phi_0 + g^{1/3} A$, 
one recovers Witten's action. The natural
ground state of the purely cubic action is $\Phi=0$. 
Since this corresponds to zero BRST
operator, it has been interpreted as a state of zero metric. 
But the purely cubic action
can be viewed as the strong coupling limit
$g\rightarrow \infty$ of (\ref{sft})\footnote{We
thank
S. Shenker for pointing this out.}. If this is similar to the strong coupling
limit of purely closed bosonic string theory,  
the natural
ground state  should be $M_{27}$. Could it be that $\Phi=0$ really
corresponds to $M_{27}$
and the fact that $Q=0$ is just the statement that there are no
open string excitations?
If so, how can one recover the metric and three-form excitations?

\vskip 1in \centerline{\bf Acknowledgements} \vskip 1cm It is a
pleasure to thank J. Brodie, S. Hellerman,  S. Shenker and S. Thomas for
discussions. G.H. thanks the Stanford theory group for its hospitality.
This work was supported in part by NSF grants
PHY-0070895 and PHY-9870115





\end{document}